\documentclass[]{spie}  

\usepackage{amsmath,amsfonts,amssymb}

\usepackage{float}
\usepackage{graphicx}
\usepackage{setspace}
\usepackage{tocloft}
\usepackage{lineno}
\usepackage[colorlinks=true, allcolors=blue]{hyperref}

\title{Freezing the speckles: focal plane wavefront sensing with the spatially-clipped self-coherent camera}

\author[a]{Joshua Liberman}
\author[c, b]{Sebastiaan Y. Haffert}
\author[b]{Jared R. Males}
\author[b]{Aditya Khandelwal}
\author[b]{Stephanie Rinaldi}
\author[d]{Joseph D. Long}
\author[a]{Katie M. Twitchell}
\author[a]{Jay K. Kueny}
\author[b]{Ewan S. Douglas}
\author[e]{Lauren Schatz}
\author[f]{Jhen Lumbres}

\affil[a]{James C. Wyant College of Optical Sciences, University of Arizona,
Meinel Building 1630 E. University Blvd., Tucson, AZ. 85721}
\affil[b]{Steward Observatory, University of Arizona, 933 N Cherry Ave, Tucson, AZ, USA 85719}
\affil[c]{Leiden University, Leiden Observatory, The Netherlands}
\affil[d]{Flatiron Center for Computational Astrophysics, Flatiron Institute, 162 5th Avenue, New York, New York, USA}
\affil[e]{Starfire Optical Range, Kirtland Air Force Base, Albuquerque, New Mexico, USA}
\affil[f]{Northrop Grumman, Pasadena, CA}

\authorinfo{Further author information: (Send correspondence to Joshua Liberman)\\Joshua Liberman: E-mail: jliberman@arizona.edu}

\begin{document} 
\maketitle

\begin{abstract}
The next generation of Extremely Large Telescopes (ELTs) and the Habitable Worlds Observatory (HWO) require active speckle suppression to directly image exo-Earths. Focal plane wavefront sensing and control allows us to detect and remove time-varying speckles through measurements of the electric field, ideally with a single-shot measurement scheme. Wavefront sensing approaches include pairwise probing (PWP) and the self-coherent camera (SCC). However, the PWP technique is time-consuming, requiring at least 4 images and reducing the speed at which aberrations can be eliminated. The classical SCC modifies a standard coronagraph design, creating a reference electric field that interferes with speckles in the final focal plane, forming Fizeau fringes. However, this design only works over small spectral bandwidths and requires significantly oversized optics, limiting its effectiveness. We demonstrate a new SCC variant, the Spatially-Clipped Self-Coherent Camera (SCSCC). The SCSCC utilizes a pinhole placed close to the Lyot stop, reducing the overall beam footprint and boosting the sensor’s spectral bandwidth by factors of 3, respectively. A beamsplitter and knife edge downstream of the Lyot stop splits the light into 2 channels: fringed and unfringed, enabling wavefront sensing with a single exposure. Time-varying speckles are frozen in place, making them easy to remove. We present the SCSCC optical design combined with the photon resolving Hamamatsu Orca-Quest 2 camera. Furthermore, we demonstrate high speed wavefront control with the SCSCC, minimizing speckle intensity by 2x within a 5-11 $\lambda / D$ dark hole region on the Comprehensive Adaptive Optics and Coronagraph Test Instrument (CACTI) at the University of Arizona. These lab tests are in preparation for the on-sky demonstration of the SCSCC with the Magellan Adaptive Optics eXtreme (MagAO-X) instrument in the Fall 2026 semester. Our results make the SCSCC a valuable wavefront sensor for upcoming missions, including the Giant Magellan Telescope and HWO.
\end{abstract}


\section{INTRODUCTION}
\label{sec:intro}  

Direct imaging is a viable method for detecting and characterizing Earth-like planets. To this end, NASA is developing the Habitable Worlds Observatory (HWO)--a space-based direct imaging telescope to identify Earth-like planets around Sun-like stars\cite{pueyo2019}. To image faint planets around bright stars, astronomers suppress starlight using a coronagraph. On-axis starlight is diffracted outward via a focal plane mask, and the diffracted beam is then blocked by a Lyot stop\cite{kenworthy_2025}. By suppressing starlight while preserving planet light, astronomers can search for atmospheric biosignatures that may indicate a planet's habitability. However, the planet-to-star flux ratio, or contrast, required for imaging Earth-like planets around Sun-like stars is $\sim 10^{-10}$ at $\approx 0.1$" separations, making the direct imaging technique difficult to implement. One of the principal limitations preventing astronomers from reaching deeper contrasts in both ground- and space-based applications is slow-varying optical aberrations, or quasi-static speckles, which cause starlight to leak through a coronagraphic system\cite{currie2023}. NASA's HWO mission, in particular, will require extremely stable optics in order to mitigate aberrations. This specification presents engineering challenges and budget constraints\cite{coyle2023, steiger2026}.

Using higher order wavefront sensing and control (HOWFSC), we can remove quasi-static speckles and reach deeper contrasts. In HOWFSC, a deformable mirror senses and subsequently minimizes the electric field in a given region of the science image. We often refer to this process as ``digging a dark hole." Model-based sensing and control techniques such as pairwise probing + electric field conjugation (PWP + EFC) have been commonly used\cite{giveon2011, groff2015}, with the aforementioned method being baselined for use in the Roman Space Telescope's coronagraphic instrument\cite{cady2025}. But these model-based HOWFSC approaches are often too slow to sense and remove temporally evolving speckles and rely on optical models that can break down on-sky\cite{haffert_implicit_2023, xin2023}. Correcting aberrated wavefronts and achieving contrasts sufficient for detecting Earth-like planets instead requires measurement-based HOWFSC techniques.

Two measurement-based wavefront sensing and control approaches are implicit electric field conjugation (iEFC) and the self-coherent camera (SCC) + iEFC. The classical iEFC technique requires at least 4 images thus reducing the speed at which aberrations can be eliminated\cite{desai2024, haffert_implicit_2023}. The SCC combines a coronagraph with a wavefront sensor (WFS), measuring the wavefront with just a single image. Furthermore, the SCC when combined with iEFC measures and controls wavefronts at least 4x faster than iEFC with PWP.

The conventional SCC utilizes a pinhole in the Lyot stop to spatially filter a portion of the light diffracted by the coronagraph. The filtered light interferes with quasi-static speckles to create fringes in the detector plane. The wavefront can then be sensed and removed using the phase information encoded within the fringes\cite{baudoz2006}. The conventional SCC has existed for 20 years but has rarely been implemented outside of a lab setting due to its narrow spectral range\cite{coyle2021, galicher_2019} and oversized optics. As such, multiple variations on the classic SCC have been developed, which utilize a broader bandwidth by moving the pinhole closer to the Lyot pupil\cite{bos2021}. This modification to the SCC's optical design is enabled by modulating the pinhole and computing a difference image between the modulated and unmodulated beams. Numerous modulated SCC designs have been proposed, however, these wavefront sensors are either limited by their modulation speed or require extensive calibrations in post-processing, limiting their efficacy on-sky\cite{martinez2019, bos2021, haffert2022}.

Our new modulated SCC flavor--the ``Spatially-Clipped SCC" (or SCSCC)--utilizes empirical measurements to sense aberrations in a single image. The concept and initial performance simulations of the SCSCC as a WFS can be found in Ref.~\citenum{liberman2025}. In this work, we design and implement the SCSCC on the Comprehensive Adaptive Optics and Coronagraph Test Instrumentation (CACTI) testbed at the University of Arizona. We also dig the first lab dark hole using the SCSCC together with iEFC for single-shot wavefront sensing and control.

In Section~\ref{sect:methods}, we detail our final optical design, lab setup, and data reduction procedure. In Section~\ref{sect:analysis}, we analyze our results, and in Section~\ref{sect:Conclusion}, we discuss and conclude our work.

\section{Methods}
\label{sect:methods}
\subsection{Design and Lab Setup}
\label{subsect: design-and-lab}
\begin{figure}[H]
\begin{center}
\begin{tabular}{c}
\includegraphics[height=6.5cm]{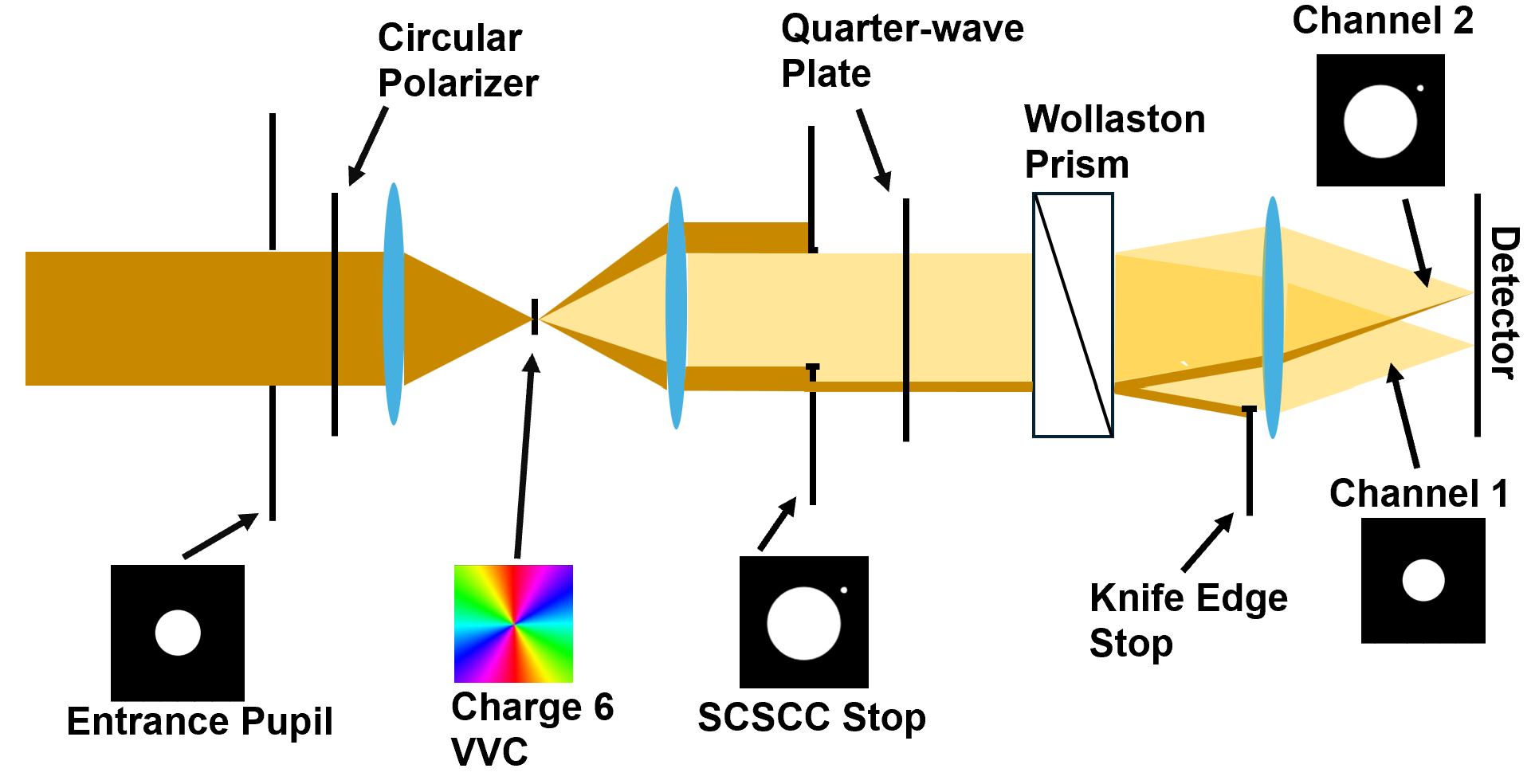}
\end{tabular}
\end{center}
\caption 
{ \label{fig:scc-design}
The SCSCC optical design. A charge 6 Vector Vortex Coronagraph (VVC) diffracts on-axis starlight (gold) outward, where it is spatially filtered by a pinhole. A Wollaston prism then splits the light into two channels. In channel 2, the spatially filtered starlight interferes with the quasi-static speckles (yellow) to create a fringed point spread function (PSF). In channel 1, the spatially filtered beam is blocked by a knife edge, producing an unfringed PSF. A circular polarizer and quarter-wave plate are placed upstream and downstream of the VVC, respectively, to remove the VVC's circular polarization leakage\cite{serabyn2019}. Our optical setup enables wavefront sensing with a single exposure.} 
\end{figure}

\begin{figure}[H]
\begin{center}
\begin{tabular}{c}
\includegraphics[height=6.5cm]{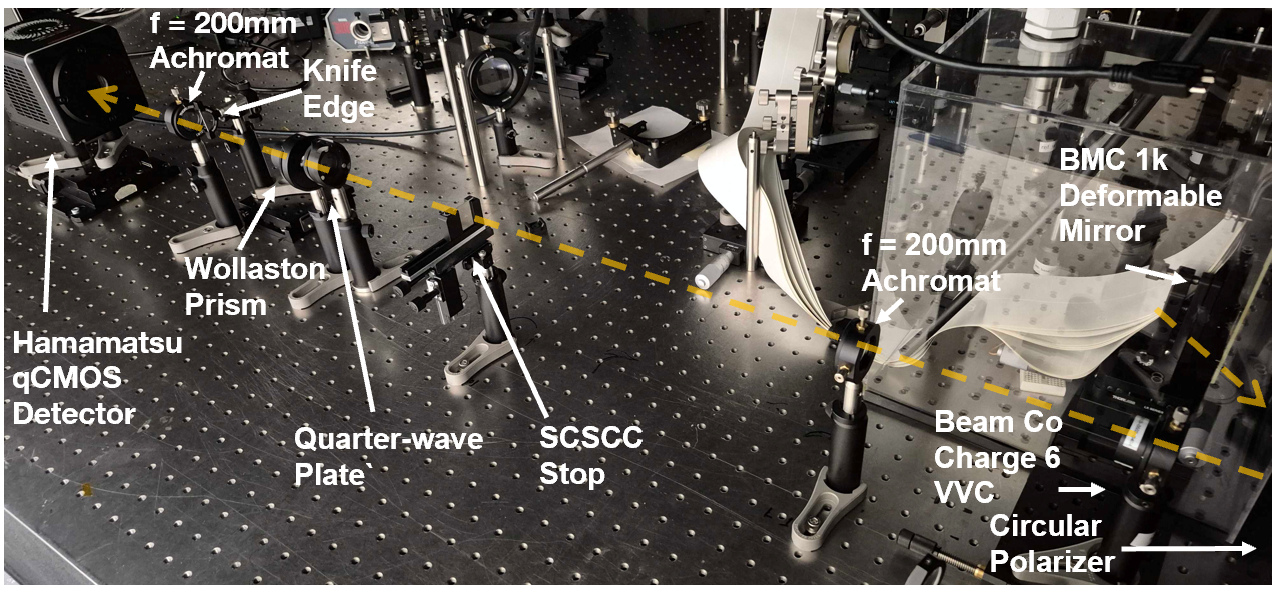}
\end{tabular}
\end{center}
\caption 
{ \label{fig:scc-lab}
The SCSCC mounted on the Comprehensive Adaptive Optics and Coronagraph Test Instrumentation (CACTI) testbed at the University of Arizona\cite{schatz2022}. The dashed orange line denotes the beam path through the optical system.} 
\end{figure}

The finalized design of our SCSCC is a modification of the initial design proposed in Ref.~\citenum{liberman2025}. In our final design, we introduce a Wollaston prism downstream of the SCSCC stop, which acts as a common-path beam splitter. In one channel, the reference beam from the pinhole is blocked by a knife edge, creating our unfringed point spread function (PSF). In the other channel, the reference beam interferes with our quasi-static speckles to create fringes in the focal plane. The choice of the Wollaston prism over a 50/50 beam splitter is motivated by the fact that we wish to minimize differential aberrations that may otherwise be introduced between the two beam paths. A schematic of this finalized SCSCC design is shown in Fig.~\ref{fig:scc-design}. In Fig.~\ref{fig:scc-lab}, we show the SCSCC as implemented on the CACTI testbed at the University of Arizona.

\begin{figure}[H]
\begin{center}
\begin{tabular}{c}
\includegraphics[height=6.5cm]{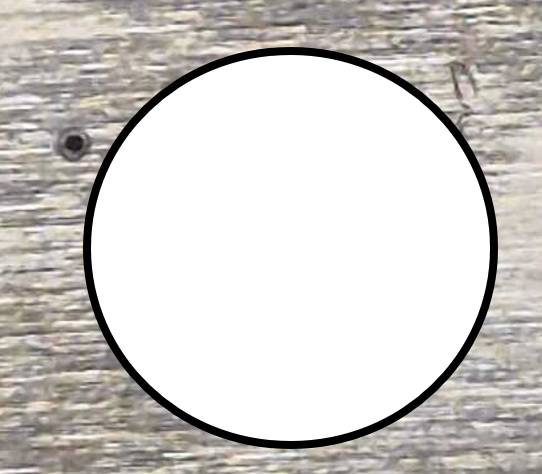}
\end{tabular}
\end{center}
\caption 
{ \label{fig:scscc-stop}
The SCSCC stop. A laser cutter is used to create the pupil stop and the pinhole on a 2.54 x 2.54 cm aluminum substrate. The stop diameter is 3.58 mm ($0.95 \times$ the entrance pupil diameter ($D_\text{EP}$) and the pinhole diameter is $71\mu m$. The separation between the stop and the pinhole is 2 mm (or $0.53 D_\text{EP}$). $0.53 D_\text{EP}$ is close to the theoretical pupil-to-pinhole separation limit where the wavefront phase information can still be recovered\cite{bos2021}.}
\end{figure}

Previously, we observed that polarization leakage from the charge 6 vector vortex coronagraph (VVC) was causing light to enter our Wollaston in a mixed circular polarization state. This effect introduced significant aberrations in our final PSFs. To mitigate aberrations caused by circular polarization leakage through the VVC, we implement polarization filtering by adding a quarter-wave plate upstream of our Wollaston prism (Fig.~\ref{fig:scc-lab}), effectively creating a circular analyzer\cite{serabyn2019}. In this configuration, we can ensure that our light is linearly polarized when entering the Wollaston prism and thus eliminate downstream aberrations caused by the polarization leakage. The SCSCC stop itself was manufactured using a laser cutter at the University of Arizona's Machining and Welding Center and is shown in Fig.~\ref{fig:scscc-stop}.

\subsection{Data Reduction}
\label{subsect:data-reduction}
\begin{figure}[H]
\begin{center}
\begin{tabular}{c}
\includegraphics[height=6.5cm]{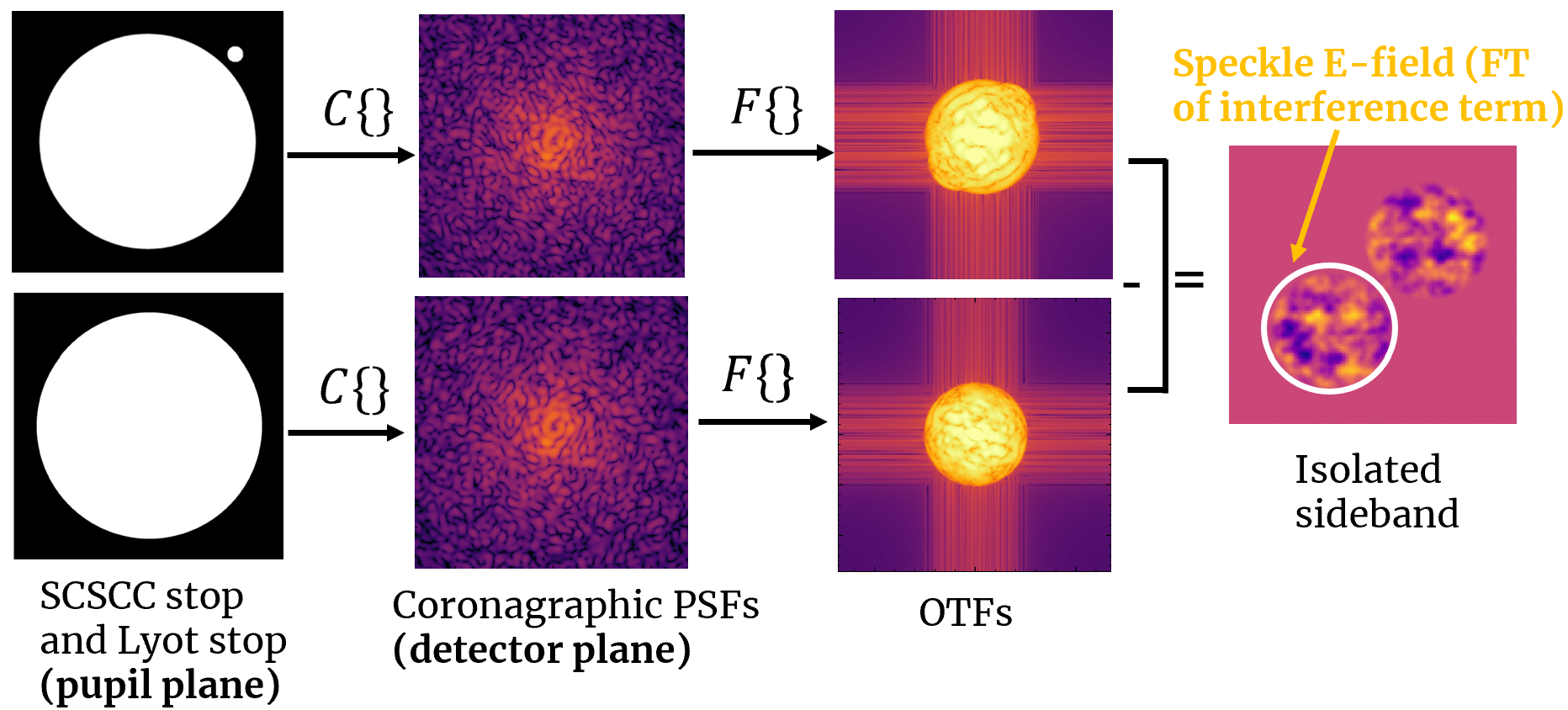}
\end{tabular}
\end{center}
\caption 
{ \label{fig:scc-mod-sims}
Simulated SCSCC images showing how the wavefront phase can be recovered. The top row depicts the wavefront propagation through our unfringed channel (channel 1), while the bottom row shows the propagation through our fringed channel (channel 2). $C\{\}$ denotes a coronagraph operator that propagates our electric field through a coronagraphic optical system, producing the coronagraphic PSF. $F\{\}$ denotes a Fourier transform operator. The real component of our OTF is shown in the figure above. The rightmost image depicts the difference between our fringed and unfringed OTFs. Subtracting our unfringed OTF from our fringed OTF, we may isolate the sidebands (the interference term between our filtered beam and the quasi-static speckles in Fourier space) that are buried within our fringed OTF. Our wavefront phase information is contained within the sidebands, where a single sideband is circled in white. Note that we only need one of the sidebands to recover our phase as the Fourier transform of our fringed PSF produces two copies of the interference term. This figure has been adapted from Ref.~\citenum{martinez2019}.}
\end{figure}

 From the fringed and unfringed PSFs obtained in channels 1 and 2, we may then recover our wavefront information by taking the difference (or `modulating') between either our fringed and unfringed optical transfer functions (OTFs) or PSFs, depending on whether we wish to perform the phase reconstruction in the Fourier plane or in the focal plane. The OTF is defined as the Fourier transform of the PSF. In the focal plane, we may use the difference image directly as the input to our sensing and control algorithm\cite{thompson2022, liberman2025}. In the Fourier plane, additional post-processing steps are required after obtaining our difference OTF. We must apply a mask to extract a single sidelobe, shift the sidelobe such that it's centered within our image, and then inverse Fourier transform the image and take the imaginary component. The imaginary component of our inverse Fourier transformed OTF contains the wavefront phase information\cite{martinez2019}. This process of isolating the sidelobes is illustrated using simulated images in Fig.~\ref{fig:scc-mod-sims}. 

\begin{figure}[H]
\begin{center}
\begin{tabular}{c}
\includegraphics[height=4cm]{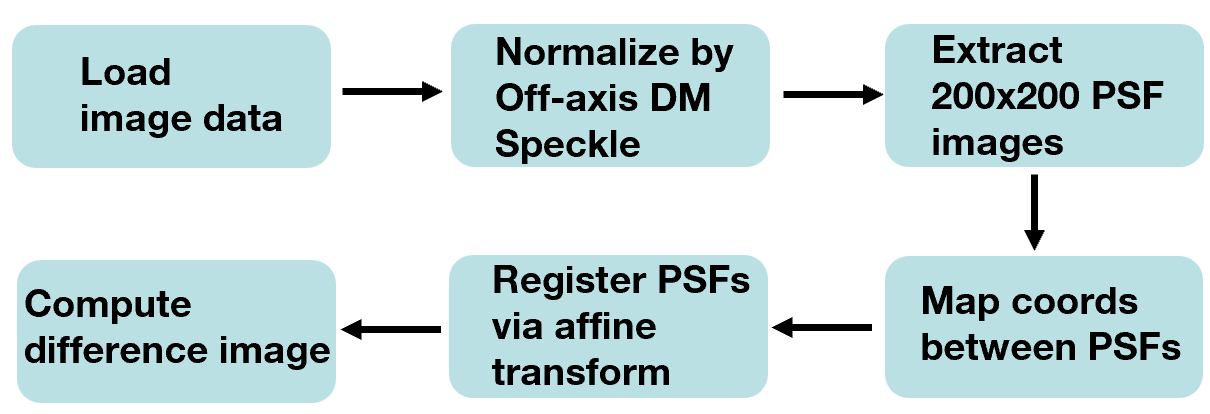}
\end{tabular}
\end{center}
\caption 
{ \label{fig:flow-diagram}
A flow diagram showing how the wavefront measurements can be acquired from experimental data. Accurately modulating the SCSCC requires sub-pixel registration between our two PSFs. We register the PSFs via an affine transform to correct for field-dependent distortion across our detector.} 
\end{figure}

\begin{figure}[H]
\begin{center}
\begin{tabular}{c}
\includegraphics[height=6.5cm]{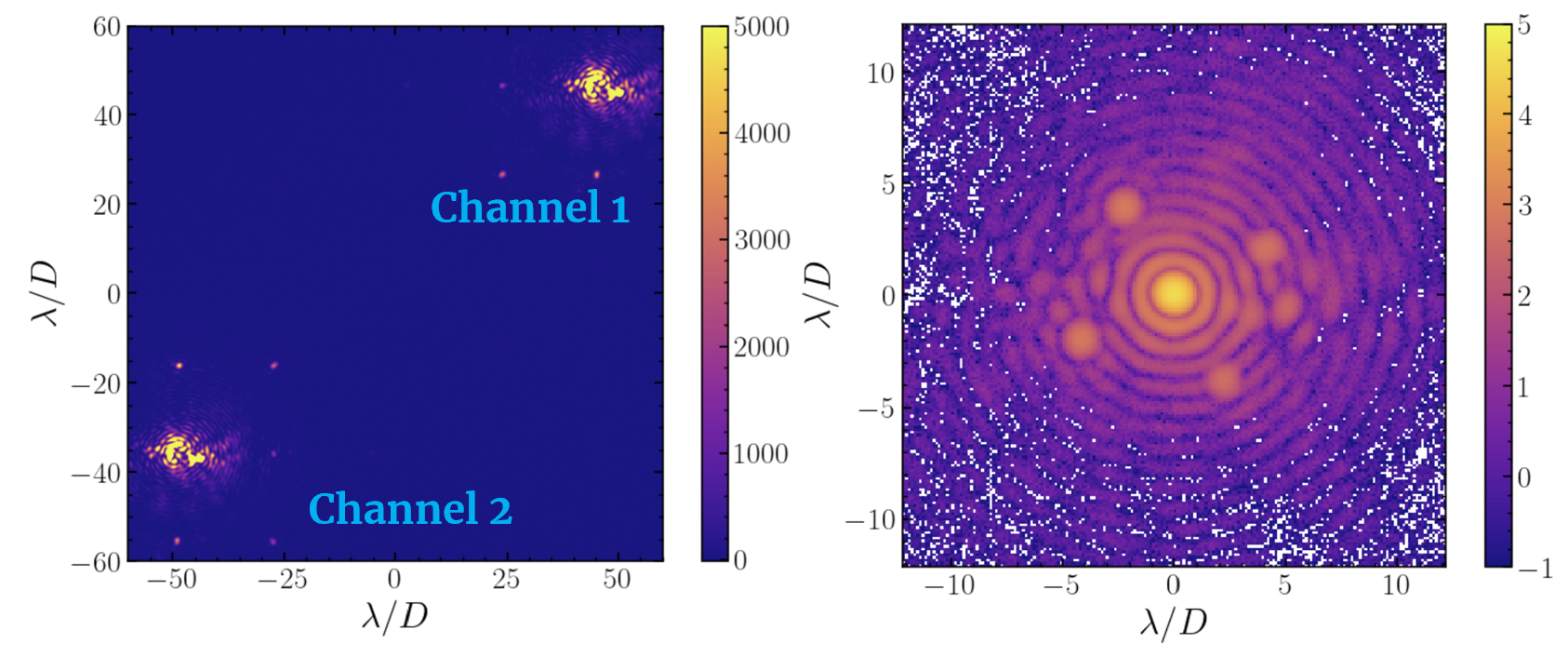}
\end{tabular}
\end{center}
\caption 
{ \label{fig:scscc-fullframe}
The full frame image showing the unfringed (channel 1) and fringed (channel 2) PSFs on the detector \textbf{(left)} and an unsaturated PSF used for creating a center-cropped region of interest, or ROI \textbf{(right)}. Four artificial deformable mirror (DM) speckles can be seen symmetrically distributed about the central PSF within the unsaturated image, The left colorbar is in units of linear scale while the right colorbar is in units of log scale.} 
\end{figure}

In Fig.~\ref{fig:flow-diagram}, we show a flow diagram that highlights the necessary steps to go from raw images to wavefront measurements. The most crucial step in our data reduction process is ensuring that our images are registered with sub-pixel accuracy. We implement our image registration using an affine transform\cite{Hartley2004} so as to mitigate differential shear between the two PSFs that is introduced by field-dependent distortions on the detector. Fig.~\ref{fig:scscc-fullframe} depicts a full frame image of the fringed and unfringed PSFs from the SCSCC at the detector plane and an unsaturated PSF. 

The data were collected in a 1200x1200 pixel region of interest (ROI) using a He-Ne laser source with $\lambda = 633$nm. The exposure time was 7.5 ms. We center-crop the individual PSFs from channels 1 and 2 around a 200x200 ROI by removing the coronagraph and applying artificial deformable mirror (DM) speckles which serve as our positional references\cite{mcewen2024, gerard2021} (Fig.~\ref{fig:scscc-fullframe} at right). Note that the PSFs shown in the following figures are not perfectly centered within the ROI. This is because the artificial DM speckle positions for the unsaturated PSF were not recorded for this particular dataset prior to re-aligning the coronagraph. As such, we used previous measurements of the artificial speckle locations that did not account for slow drifts within the testbed. The fluxes from the two PSFs are normalized with respect to one another, using the intensity ratio between corresponding DM diffraction speckles from channels 1 and 2 as a scaling factor. These diffraction speckles are caused by the periodic actuator structure behind the DM face sheet\cite{twitchell2026} and can be seen evenly distributed around the PSFs in Fig.~\ref{fig:scscc-fullframe} (left).

\begin{figure}[H]
\begin{center}
\begin{tabular}{c}
\includegraphics[height=6.5cm]{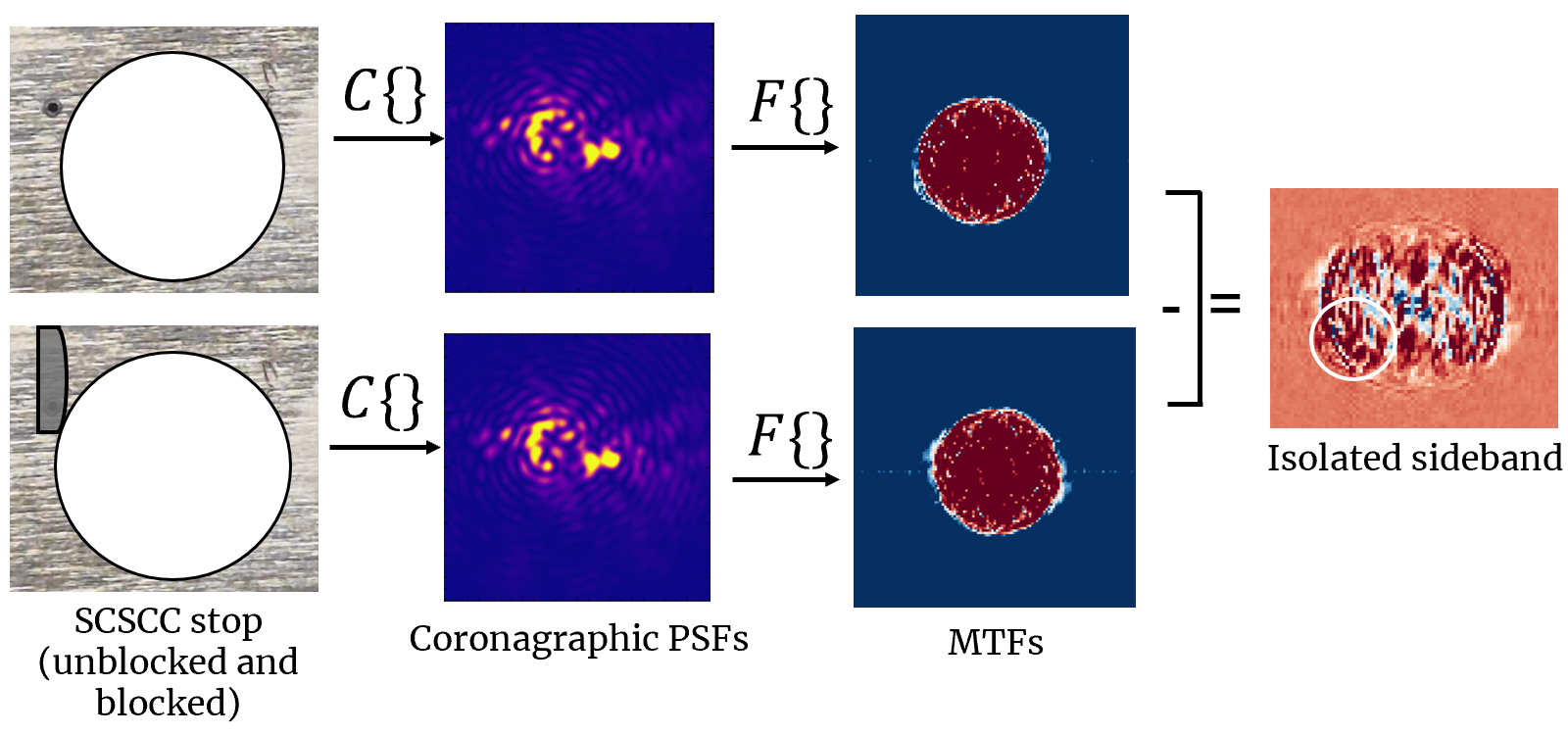}
\end{tabular}
\end{center}
\caption 
{ \label{fig:wf-recovery}
Lab data showing our ability to recover the speckle electric field in the Fourier plane. For illustrative purposes, the modulation transfer function (MTF), defined as the absolute magnitude of our OTF, is plotted for each of the channels. Channel 2 \textbf{(top)} denotes our fringed data while channel 1 \textbf{(bottom)} denotes our unfringed data. The pinhole in channel 1 is blocked by a knife edge. An isolated sideband after subtracting our unfringed MTF from our fringed MTF is circled in white.} 
\end{figure}

We then create a point-to-point coordinate mapping between our two center-cropped PSFs from channels 1 and 2, respectively. We do so by identifying seven speckle features that are common between our channel 1 (source) and channel 2 (destination) PSFs and feeding the corresponding source and destination input coordinates as a transformation matrix into OpenCV's \texttt{warpaffine} function\cite{opencvlibrary}. After registering our PSFs in channels 1 and 2, we can recover our phase information using difference images. This procedure is shown in Fig.~\ref{fig:wf-recovery}.

\begin{figure}[H]
\begin{center}
\begin{tabular}{c}
\includegraphics[height=6.5cm]{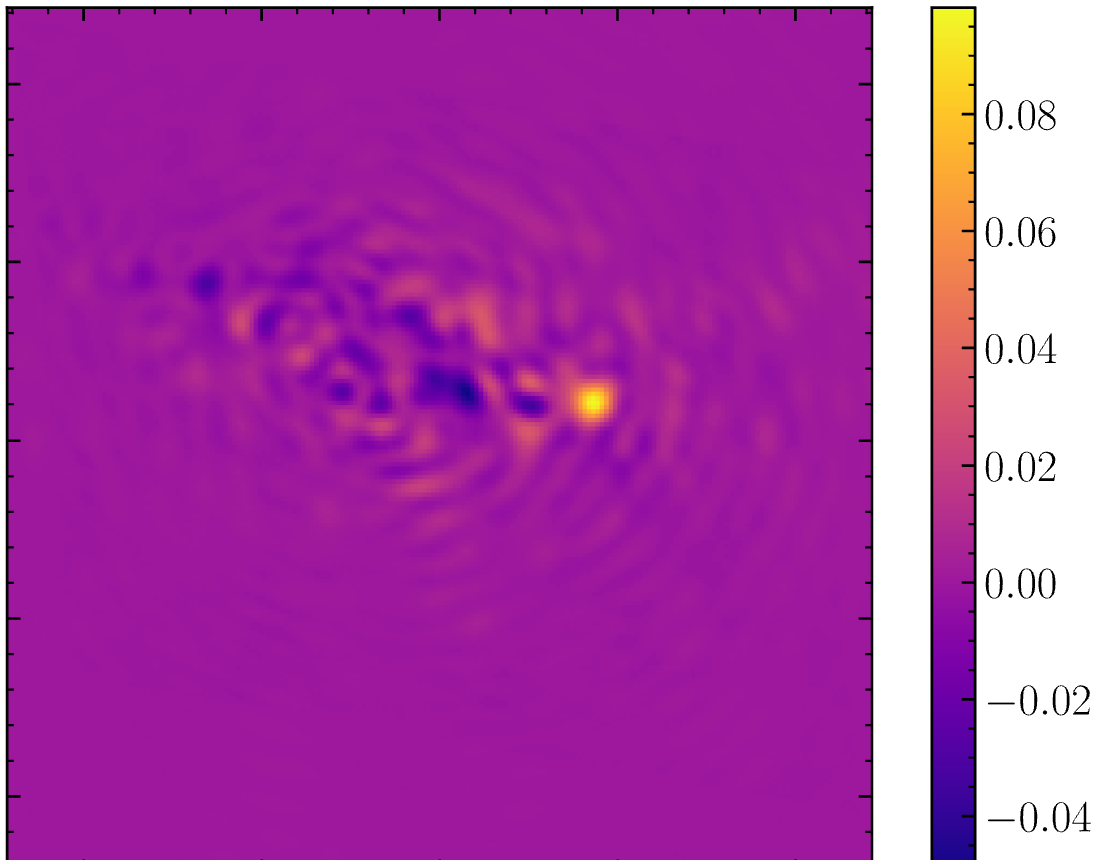}
\end{tabular}
\end{center}
\caption 
{ \label{fig:diff-psf}
A focal plane difference image between the registered PSFs of channels 1 and 2. The largest residuals are distributed along the x-axis and may be caused by constructive interference between the sinusoidal fringe pattern and the quasi-static speckles. The colorbar is expressed in units of pixel counts.} 
\end{figure}

In our channel 2 PSF shown in Fig.~\ref{fig:wf-recovery} (top), the fringes are not visible. The low fringe visibility is caused by the small pinhole diameter relative to that of the stop (see Fig.~\ref{fig:scscc-stop} for the stop specifications). The thickness of the knife edge also introduces some diffraction artifacts which can be seen around the central lobe of the channel 1 modulation transfer function (MTF) in Fig.~\ref{fig:wf-recovery} (bottom). These diffraction artifacts likely contribute to the residuals that we observe in our difference MTF. Additional residuals are likely caused by inaccuracies in our image registration between channels 1 and 2. Despite the imperfect MTF subtraction, we are still able to identify the side lobes, which are circled in white in Fig.~\ref{fig:wf-recovery}. Moving forward, we elect to use the PSF difference images as the wavefront measurements within our iEFC loop due to ease of implementation. One such difference PSF is shown in Fig.~\ref{fig:diff-psf}.

Using focal plane difference images, we may empirically calibrate an interaction matrix relating DM modal coefficients to WFS measurements. The detailed mathematical formalism behind the SCSCC can be found in Ref.~\citenum{liberman2025}. Below, we summarize the SCSCC calibration procedure in the following steps:

\begin{enumerate}
    \item Construct an interaction matrix relating DM modal coefficients to WFS measurements
        \begin{enumerate}
            \item Apply Fourier mode on DM
            \item Capture difference image (fringed - unfringed) in focal plane
            \item Repeat for all modes contained in the basis
        \end{enumerate}
    \item Invert matrix with Tikhonov regularization
    \item Multiply difference (measurement) image by inverted matrix to obtain DM modal coefficients
\end{enumerate}
The DM solution that minimizes intensity can then be expressed as

\begin{equation}
    a = \arg\!\min_{a} |\Delta I + (\textbf{H}a)|^{2} + \lambda|a|^{2}
\end{equation}
where $a$ is our DM solution, $\Delta I$ is a wavefront measurement, \textbf{H} is our interaction matrix with shape $N_\text{modes} \times M_{\text{pixels}}$, and $\lambda$ is a regularization coefficient for penalizing solutions that contain large actuator stroke. The integrated sensing and control procedure above can be described as SCSCC+iEFC, where we implicitly minimize intensity difference images from the SCSCC.

\section{Analysis}
\label{sect:analysis}
\begin{figure}[H]
\begin{center}
\begin{tabular}{c}
\includegraphics[height=6.5cm]{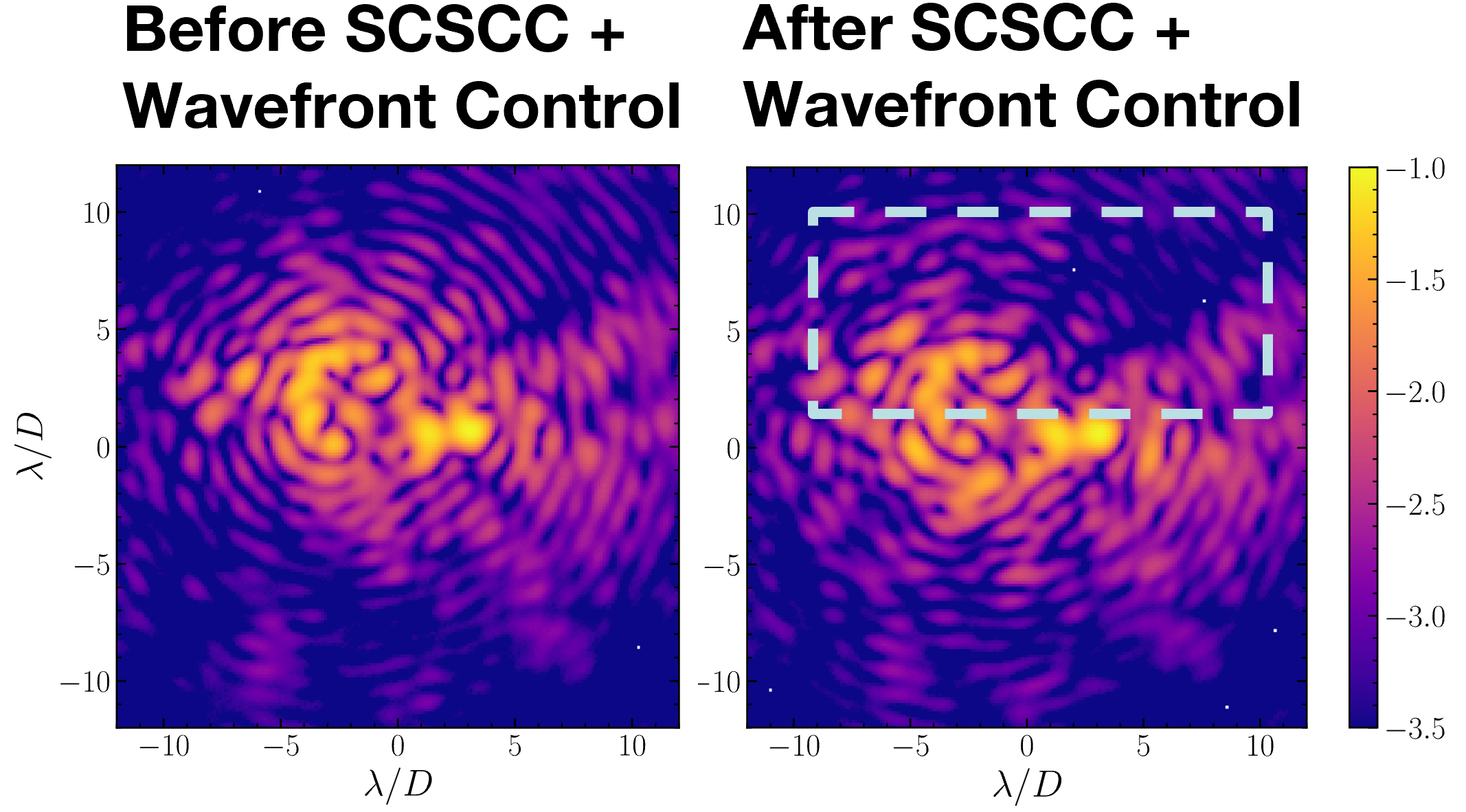}
\end{tabular}
\end{center}
\caption 
{ \label{fig:lab-dh}
Post-coronagraphic PSFs before \textbf{(left)} and after \textbf{(right)} wavefront control. The control region is highlighted with a dashed blue square. The colorbar denotes normalized intensity in log scale. We define normalized intensity as the measured intensity in the post-coronagraphic PSF divided by the measured intensity of an off-axis DM speckle.} 
\end{figure}

We dig a dark hole in lab, specifying a control region with an inner working angle of 2$\lambda / D$, an outer working angle of 10 $\lambda / D$, and a width of 20 $\lambda / D$. Our control region contains 360 Fourier modes and our matrix is calibrated with a mode amplitude of 0.01 $\lambda$. Our dark hole results are shown in Fig.~\ref{fig:lab-dh}.

We are only able to minimize speckle intensity in less than half of the control region, which may be due to low order aberrations limiting our starting contrast. We also find that our PSF registration accuracy varies across calibration images. These inaccuracies may be caused by non-uniformities between the channel 1 and channel 2 PSFs that are due to field-dependent distortion. While we attempt to correct for this distortion via an affine transform, we assume a fixed image structure when performing our point-to-point coordinate mapping. However, the speckle field varies with every Fourier mode applied on the DM, causing the registration to exhibit inconsistencies between calibration frames.

\begin{figure}[H]
\begin{center}
\begin{tabular}{c}
\includegraphics[height=6.5cm]{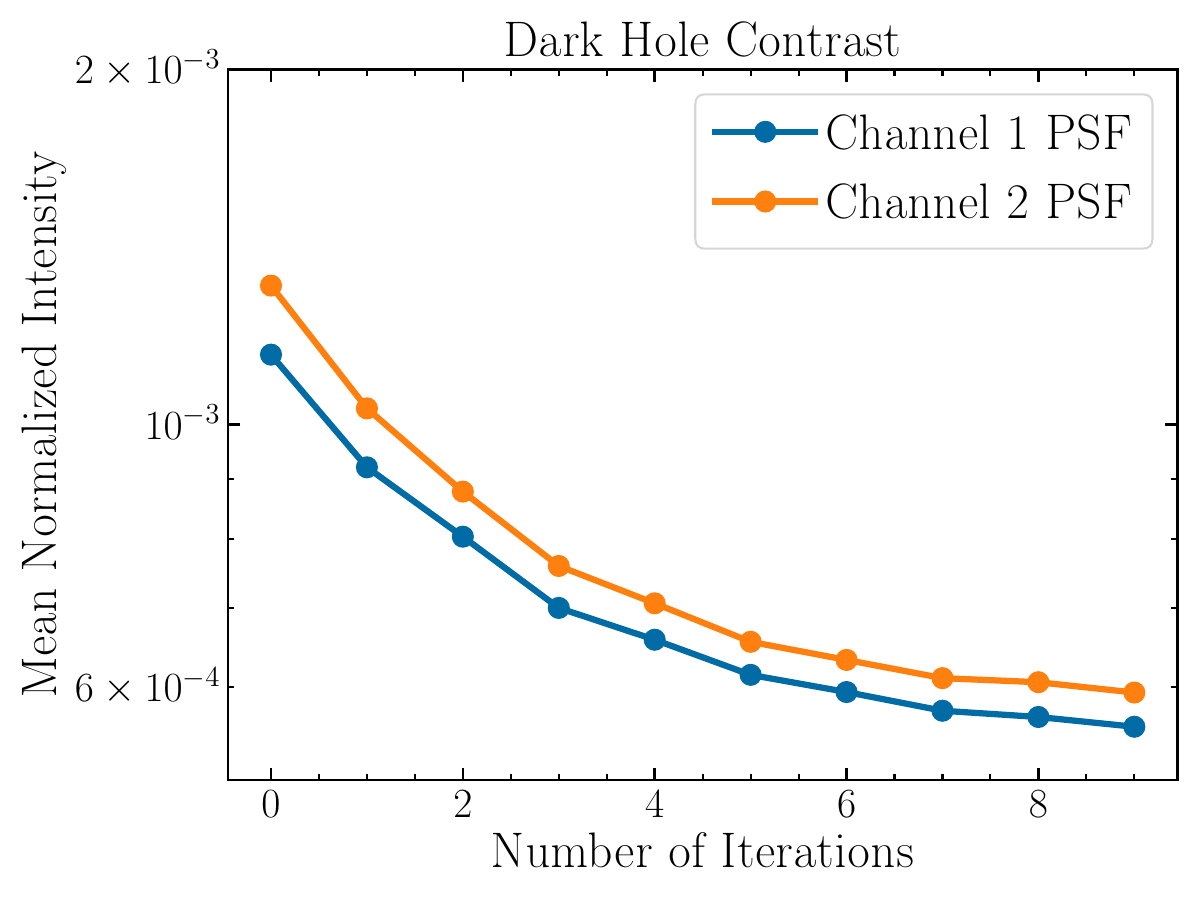}
\end{tabular}
\end{center}
\caption 
{ \label{fig:contrast-curve}
A contrast curve showing the mean normalized intensity over 10 iterations of wavefront control. While our specified control region is from 2-10 $\lambda / D$ with a width of 20 $\lambda / D$, we only measure the normalized intensity over a 5-10 $\lambda / D$ rectangular area and a width of 8 $\lambda / D$. This is because the speckles in the left half of our specified control region are poorly sensed.} 
\end{figure}

Our final contrast curve is shown in Fig.~\ref{fig:contrast-curve} with a measurement region that spans 5-10 $\lambda / D$ with a width of 8 $\lambda / D$. We use a measurement region that is considerably smaller than our defined control region, as the speckles in the left half of our specified control region are poorly sensed. We observe a reduction in speckle intensity of $\approx 2\times$ within our measurement region, with a mean normalized intensity going from $1\times10^{-3}$ to $5\times10^{-4}$ after 10 iterations of wavefront control. We also observe subtle intensity differences between our channel 1 and channel 2 PSFs which may be due to polarization aberrations through the Wollaston prism.

\section{Conclusion}
\label{sect:Conclusion}
In this work, we dig the first dark hole with our combined SCSCC+iEFC sensing and control method on the CACTI testbed at the University of Arizona. We demonstrate that it is possible to modulate the SCC using a single image, allowing us to efficiently remove temporally evolving speckles. Additionally, the SCSCC has both a higher spectral bandwidth and a reduced beam footprint relative to that of the classical SCC. We are able to minimize speckle intensity by a factor of 2 within a 5-10 $\lambda / D$ dark hole. However, this dark hole only covers about one third of our specified control region. We expect both our reduction in contrast and the uniformity of our dark hole to improve significantly by increasing the accuracy of our PSF registration. 

Future work will involve refining the data reduction pipeline so as to sense the wavefront at higher SNR. One solution to the PSF registration inaccuracies is to fit for the ideal image transformation parameters that minimize residuals between the source and destination images, using a least-squares minimization approach. We also plan to utilize focus diversity phase retrieval on the CACTI testbed to remove residual low order aberrations that are present in our PSF\cite{vangorkom2021, kueny2024}. These aberrations result in bright speckles leaking through the VVC and limiting our starting contrast. Finally, in November, 2026, we will be demonstrating the SCSCC on-sky with the MagAO-X\cite{males2024} instrument.

\acknowledgments
The authors would like to thank Pierre Baudoz, Axel Potier, Scott Will, and Ben Gerard for helpful discussions regarding how best to improve the SCSCC design and data reduction procedure. J.L. and S.Y.H acknowledge support from NASA APRA grant \#80NSSC24K0288. This research made use of community-developed core Python packages, including: HCIPy\cite{por2018}, Astropy\cite{robitaille2013}, Matplotlib\cite{hunter2007}, and the IPython Interactive Computing architecture\cite{perez2007}. Joshua Liberman is a member of UCWAZ Local 7065.
\bibliography{report} 
\bibliographystyle{spiebib} 

\end{document}